\documentclass[aps,floatfix,prd,showpacs]{revtex4}
\usepackage{graphicx}
\usepackage{dcolumn}
\usepackage{bm}
\usepackage{amsmath,amsthm,amssymb,epsfig,alltt}

\begin{document}

\def\a{\alpha}
\def\b{\beta}
\def\d{{\delta}}
\def\l{\lambda}
\def\e{\epsilon}
\def\p{\partial}
\def\m{\mu}
\def\n{\nu}
\def\t{\tau}
\def\th{\theta}
\def\s{\sigma}
\def\g{\gamma}
\def\G{\Gamma}
\def\o{\omega}
\def\r{\rho}
\def\z{\zeta}
\def\D{\Delta}
\def\half{\frac{1}{2}}
\def\hatt{{\hat t}}
\def\hatx{{\hat x}}
\def\hatp{{\hat p}}
\def\hatX{{\hat X}}
\def\hatY{{\hat Y}}
\def\hatP{{\hat P}}
\def\haty{{\hat y}}
\def\whatX{{\widehat{X}}}
\def\whata{{\widehat{\alpha}}}
\def\whatb{{\widehat{\beta}}}
\def\whatV{{\widehat{V}}}
\def\hatth{{\hat \theta}}
\def\hatta{{\hat \tau}}
\def\hatrh{{\hat \rho}}
\def\hatva{{\hat \varphi}}
\def\barx{{\bar x}}
\def\bary{{\bar y}}
\def\barz{{\bar z}}
\def\baro{{\bar \omega}}
\def\barpsi{{\bar \psi}}
\def\sp{\sigma^\prime}
\def\nn{\nonumber}
\def\cb{{\cal B}}
\def\2pap{2\pi\alpha^\prime}
\def\wideA{\widehat{A}}
\def\wideF{\widehat{F}}
\def\beq{\begin{eqnarray}}
 \def\eeq{\end{eqnarray}}
 \def\4pap{4\pi\a^\prime}
 \def\xp{{x^\prime}}
 \def\sp{{\s^\prime}}
 \def\ap{{\a^\prime}}
 \def\tp{{\t^\prime}}
 \def\zp{{z^\prime}}
 \def\xpp{x^{\prime\prime}}
 \def\xppp{x^{\prime\prime\prime}}
 \def\barxp{{\bar x}^\prime}
 \def\barxpp{{\bar x}^{\prime\prime}}
 \def\barxppp{{\bar x}^{\prime\prime\prime}}
 \def\barchi{{\bar \chi}}
 \def\baro{{\bar \omega}}
 \def\bpsi{{\bar \psi}}
 \def\barg{{\bar g}}
 \def\barz{{\bar z}}
 \def\bareta{{\bar \eta}}
 \def\ta{{\tilde \a}}
 \def\tb{{\tilde \b}}
 \def\tc{{\tilde c}}
 \def\tz{{\tilde z}}
 \def\tJ{{\tilde J}}
 \def\tpsi{\tilde{\psi}}
 \def\tal{{\tilde \alpha}}
 \def\tbe{{\tilde \beta}}
 \def\tga{{\tilde \gamma}}
 \def\tchi{{\tilde{\chi}}}
 \def\barth{{\bar \theta}}
 \def\bareta{{\bar \eta}}
 \def\barom{{\bar \omega}}
 \def\bole{{\boldsymbol \epsilon}}
 \def\bolth{{\boldsymbol \theta}}
 \def\bomega{{\boldsymbol \omega}}
 \def\bolmu{{\boldsymbol \mu}}
 \def\bola{{\boldsymbol \alpha}}
 \def\bolb{{\boldsymbol \beta}}
 \def\bolX{{\boldsymbol X}}
 \def\mathN{{\boldsymbol n}}
 \def\bba{{\boldsymbol a}}
 \def\bby{{\boldsymbol y}}
 \def\bbp{{\boldsymbol p}}
 \def\bbA{{\boldsymbol A}}
 \def\bbphi{{\boldsymbol \phi}}
 \def\mathP{{\mathbb P}}
 \def\mathN{{\boldsymbol N}}
 \def\mathN{{\mathbb N}}
 \def\bbP{{\boldsymbol P}}


\title{Four-Gauge-Particle Scattering Amplitudes and \\
Polyakov String Path Integral in the proper-time gauge}

\author{Taejin Lee}
\affiliation{
Department of Physics, Kangwon National University, Chuncheon 24341
Korea}

\email{taejin@kangwon.ac.kr}

\begin{abstract}
We evaluate four-gauge-particle tree level scattering amplitudes 
using the Polyakov string path integral in the proper-time gauge, where the string path integral can be cast into the Feynman-Schwinger proper-time representation. We compare the resultant scattering amplitudes, which include $\ap$-corrections, with the conventional ones that may be obtained by substituting local vertex operators for the external string states. In the zero-slope limit, both amplitudes are reduced to the four-gauge-particle scattering amplitude of non-Abelian Yang-Mills gauge theory. However, when the string corrections become relevant with finite $\ap$, 
the scattering amplitude in the proper-time gauge differs from the conventional one: 
The Polyakov string path integral in the proper-time gauge, equivalent to the deformed cubic string field theory, systematically provides the alpha prime corrections. In addition, we 
find that the scattering amplitude in the proper-time gauge contains tachyon poles in a manner consistent with three-particle-scattering amplitudes. The scattering amplitudes evaluated 
using the Polyakov string path integral in the proper-time gauge may be more suitable than conventional ones for exploring string corrections to the quantum field theories and high energy behaviors of open string.
\end{abstract}




\maketitle

\section{Introduction}

Scattering amplitudes, which have yielded many important new discoveries, have long been ubiquitous tools in both experimental and theoretical physics. In string theory, theoretical studies of string scattering amplitudes are also expected to lead to new findings in high energy physics, where string theory is considered 
the most promising candidate for a unified framework of the fundamental forces, including gravity. In some scenarios 
\cite{Lykken1996,Antoniadis1998} involving embedding the standard model in the frame
work of string theory, the string scale may be as low as the weak scale. These theoretical proposals allow for
new possibilities that we may directly study string physics at high energy colliders. Therefore, it becomes an important and urgent task to accurately calculate the multi-particle scattering amplitudes in string theory.

Conventionally, the string scattering amplitudes are calculated by substituting the local vertex operators for external string states. This procedure is based on the one-to-one correspondence between string states and local operators \cite{Polchinski.book.1998}. This method of calculation for string scattering amplitude has proven
successful for producing the four-tachyon scattering amplitudes 
known as the Veneziano amplitude \cite{Veneziano68} for open string and the Virasoro-Shapiro amplitude
\cite{Virasoro69,Shapiro69} for closed string. However, it is not clear whether we can apply this vertex operator technique to more general cases, such as evaluating high energy scattering amplitudes or alpha-prime corrections. In fact, the validity of this procedure has never been examined thoroughly.

An alternative method to evaluate the string scattering amplitudes, which does not make use of vertex operators, is the Polyakov string path integral \cite{Polyakov1981}. By evaluating the Polyakov string path integral defined on the string world-sheet with appropriately chosen boundary conditions for the external string states, we may obtain the string scattering amplitudes. Although this technique is more complicated than the vertex operator technique, it offers a number of advantages: 
The external string states do not need to be on-shell and the momenta of external strings may not be restricted to the low energy region. If we choose the proper-time gauge \cite{TLee88ann} in
fixing the reparametrization invariance on the string world-sheet covariantly, we can
cast string scattering amplitudes into those of second quantized theory in a fashion similar to the Feynman-Schwinger representation of quantum field theory. The string field theory \cite{Lee2016i,TLee2017cov} defined by the Polyakov string path integral in the proper-time gauge has been shown to be equivalent to the deformed cubic string field theory \cite{Lee2017d,Lai2017S}.

As for the four-tachyon-scattering amplitudes, both methods yield the same result:
The well-known Veneziano amplitude and the Virasoro-Shapiro amplitude. However, a recent work \cite{Lai2017S}
has pointed out that the two methods may produce different results if we apply them to more general external string states. A simple extension of the Veneziano amplitude is the scattering amplitude of three tachyons and one arbitrary string state, which has been studied extensively in Refs. \cite{Chan2006notes,Lai2016string,Chan2005solving,Lai2016lauri,Lai2017sloving} to explore symmetric properties of the string scattering amplitudes in the high energy limit \cite{Gross87,Gross88,Gross88prl,Gross89Phil,Gross89nucl}. When calculating the string scattering amplitude defined by the Polyakov string path integral, we must
map the string world-sheet onto upper half complex plane by the Schwarz-Christoffel transformation. On the other hand, if we adopt the conventional vertex operator technique, the string scattering amplitude is readily defined on upper-half complex plane. It follows then that the two methods may yield the same result only when the scattering amplitude is invariant under the conformal transformation generated by the Schwarz-Christoffel mapping.

In this present work, we shall study the four-gauge-particle tree level scattering amplitude in bosonic open string theory
explicitly evaluating the Polyakov string path integral in the proper-time gauge and compare the resulant scattering amplitude with the conventional expression obtained by the vertex operator technique. We will examine their high energy behaviors and the difference of the singularity structures.

\vskip 0.5cm
\begin{figure}[htbp]
   \begin {center}
    \epsfxsize=0.4\hsize

	\epsfbox{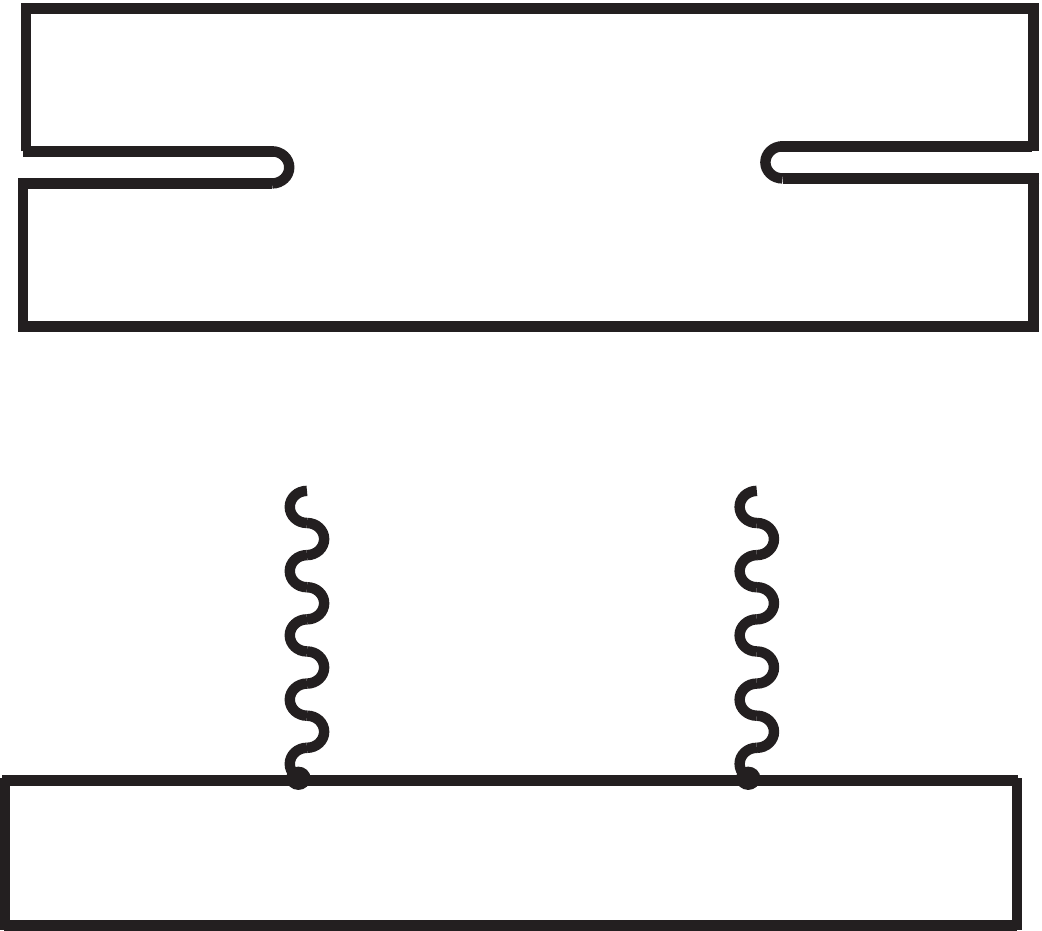}
   \end {center}
   \caption {\label{threescatopen} Four-open string scattering amplitudes. Two external strings are replaced by local vertex operators.}
\end{figure}

\section{Open String Fields on multiple space-filling Branes}

The multiple string scattering amplitude may be written in terms of the Polyakov string path integral defined on the corresponding string world-sheet \cite{GreenSW1987}
\begin{subequations}
\beq
{\cal A}_{[N]} &=& \int D[X]D[h] \exp \left(iS + i \int_{\p M} \sum_{r=1}^N  P^{(r)}\cdot X^{(r)} d\s \right), \label{A4}\\
S &=& -\frac{1}{4\pi} \int_M d\t d\s \sqrt{-h} h^{IJ} \frac{\p X^\m}{\p \s^I} \frac{\p X^\n}{\p \s^J} \eta_{\m\n}, ~~~~~ I, J = 0, \dots , d-1  
\eeq
\end{subequations}
where $\s^1= \t$, $\s^2 = \s$ and $d=26$ for open bosonic string and $d=10$ for open super-string. On a space-filling brane, the string coordinates $X^I$, satisfying the Neumann boundary condition
\beq
\frac{\p X^I}{\p \s} \Bigl\vert_{\s=0, \pi} &=& 0, ~~~\text{for} ~~I = 0, 1, \dots, d-1 , 
\eeq 
may be expanded in terms of normal modes as
\beq
X^I (\t, \s) &=& x^\m(\t) + 2 \sum_{n=1} \frac{1}{\sqrt{n}} x^\m_n(\t) \cos \left(n \s\right), ~~~ I = 0, 1, \dots, d-1. \label{xmodem}
\eeq 
If we choose the proper-time gauge where the proper-time on the string world-sheet is defined properly \cite{TLee88ann},
\beq
\p_\t N_{10} = 0, ~~~ N_{1n} = 0, ~~~ N_{2n} = 0, ~~~ n\not=0,
\eeq 
we may recast the string scattering amplitudes ${\cal A}_{[N]}$ into the Feynman-Schwinger proper-time 
representaion, which may help obtain a covariant second quantized string theory. 
Here $N_{\a n}$, $\a=1, 2$ are normal modes of the lapse and shift functions 
$N_\a = \sum_n N_{\a n} e^{in\s}$ of the two-dimensional metric on the 
world-sheet 
\beq
\sqrt{-h}h^{\a\b} = \frac{1}{N_1} \left(\begin{array}{rc} - 1  & N_2 \\ N_2 & (N_1)^2 -(N_2)^2 \end{array} \right).
\eeq 

Evaluating ${\cal A}_{[2]}$ defined as the Polyakov path integral over a strip, we can obtain the covariant 
free string propagator of the open string. The Polyakov string path inegral, ${\cal A}_{[3]}$ in the proper-time gauge has been calculated in Refs. \cite{Lee2016i,Lee2017d} and the three-string scattering amplitude ${\cal I}_{[3]}$ has been found to be
\begin{subequations}
\beq
{\cal I}_{[3]} &=& \frac{2g}{3}\int \prod_{r=1}^3 dp^{(r)} \d \left(\sum_{r=1}^3 p^{(r)}\right) \, 
\text{tr}\, \langle \Psi^{(1)}, \Psi^{(2)}, \Psi^{(3)} \vert \exp \Bigl\{ E_{[3]}[1,2,3] \Bigr\}\vert 0; p \rangle, \label{cubicterm} \\
E[1,2,3] &=& \half \sum_{n,m =1}^\infty\sum_{r,s=1}^3 \bar N^{rs}_{nm} \a^{(r)}_{-n} \cdot \a^{(s)}_{-m} + 
\sum_{n =1}^\infty \sum_{r=1}^3\bar N^r _n \a^{(r)}_{-n} \cdot \bbP +  \t_0\sum_{r=1}^3 \frac{1}{\a_r} \left(\frac{(p^{(r)})^2}{2}-1 \right), \label{E123}
\eeq 
\end{subequations}
where $\bbP = p^{(2)}-p^{(1)}$ and $\vert \Psi^{(r)}\rangle$, $r=1,2,3$ denote the external string states which carry $U(N)$ group indices. For the proper-time gauge, 
\beq
\a_1=1, \a_2=1, \a_3 =-2, \,\text{and}~ \t_0 = -2\ln 2 .  
\eeq 
Expicit expressions of the Neumann functions for the three-string scattering, $\bar N^{rs}_{nm}$ and $\bar N^r _n$ can be found in Ref. \cite{Lee2016i}:
\begin{subequations}
\beq
\bar N^{11}_{11} &=& \frac{1}{2^4}, ~~~ \bar N^{22}_{11} = \frac{1}{2^4}, ~~~ \bar N^{33}_{11} = 2^2, \label{neumanna}\\
\bar N^{12}_{11} &=& \bar N^{21}_{11} = \frac{1}{2^4}, ~~~ \bar N^{23}_{11} = \bar N^{32}_{11} = \half, 
~~~ \bar N^{31}_{11} = \bar N^{13}_{11} = \half ,\label{neumannb}\\
\bar N^1_1 &=& \bar N^2_1 = \frac{1}{4}, ~~~ \bar N^3_1 = -1 . \label{neumannc}
\eeq
\end{subequations}

If we expand the external string states in terms of mass eigen-states, 
\beq
\vert  \Psi^{(r)} \rangle &=& \phi(p^{(r)})\vert 0 \rangle + A_\m(p^{(r)}) a_1^\m \vert 0 \rangle + 
\cdots ,
\eeq 
we may obtain various three-particle interaction terms. Fig. \ref{threescatopen} depicts an expansion of three-string scattering into those of various three-particle scatterings. 

\vskip 0.5cm
\begin{figure}[htbp]
   \begin {center}
    \epsfxsize=0.8\hsize

	\epsfbox{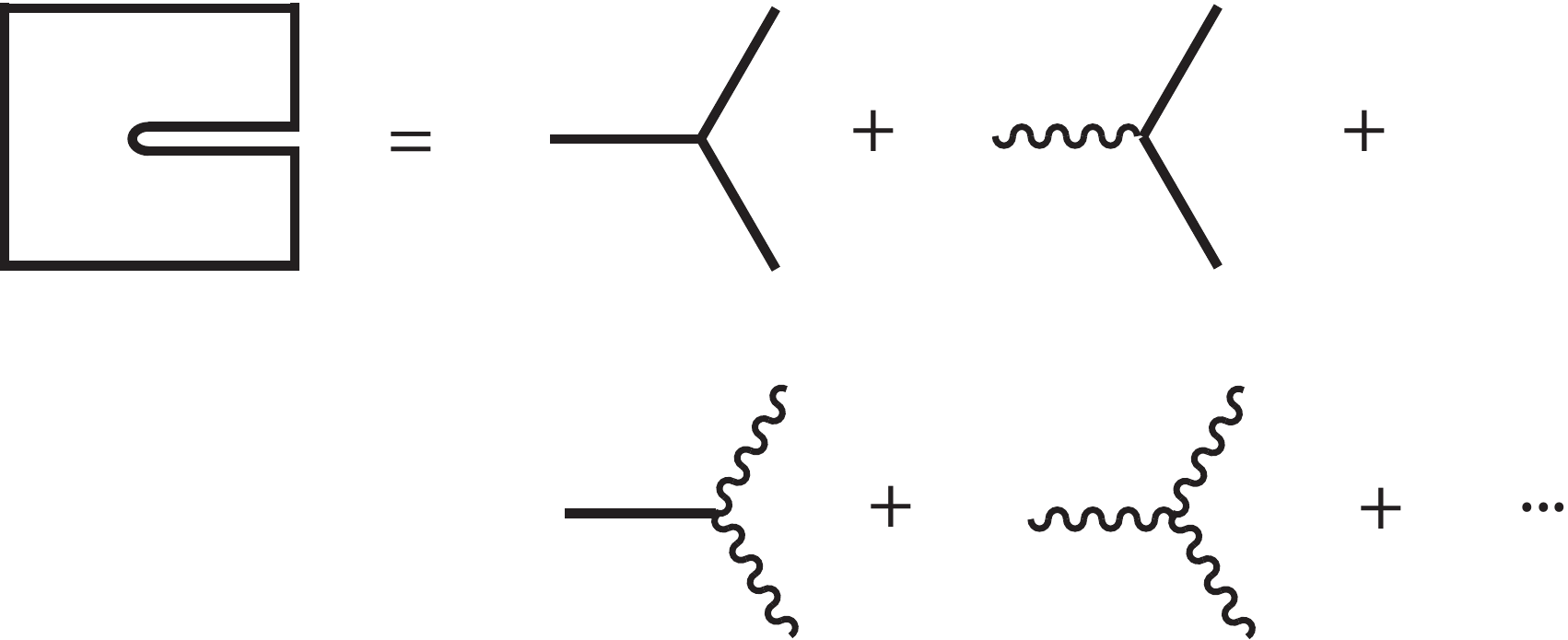}
   \end {center}
   \caption {\label{threescatopen} Three-open string scattering and three-particle scattering amplitudes.}
\end{figure}

By choosing $\prod_r \phi^{(r)} \vert 0; p \rangle$ as the external three-string state, we 
obtain the three-tachyon interaction from Eq. (\ref{cubicterm})
\beq
{\cal I}_{\phi\phi\phi} = \frac{2g}{3}\int \prod_{r=1}^3 dp^{(r)} \d \left(\sum_{r=1}^3 p^{(r)}\right) \, 
\text{tr}\, \phi(p^{(1)})\phi(p^{(2)}) \phi(p^{(3)}) . 
\eeq  
If we are interested in the three-particle interaction terms between tachyon and gauge particle, we may 
choose the external three-string external state as follows
\beq
\langle \Psi^{(1)}, \Psi^{(2)}, \Psi^{(3)} \vert = \langle 0 \vert \prod_r \left(\bbphi(r) + \bbA(r)\right)
\eeq 
where $\bbphi(r) = \phi(p^{(r)}), ~\bbA(r) = A_\m(p^{(r)}) a^{(r)\m}_{1}, ~ r= 1, 2, 3.$
Expanding the external string state in terms of component fields, from Eq. (\ref{cubicterm}) we may get three-particle interactions between tachyons and gauge particles 
\begin{subequations}
\beq
{\cal I}_{\phi\phi A} &=& \frac{2g}{3}\int \prod_{r=1}^3 dp^{(r)} \d \left(\sum_{r=1}^3 p^{(r)}\right) \, 
\text{tr}\, \langle 0 \vert \bigl(\bbphi(1)\bbphi(2) \bbA(3) + \bbphi(1)\bbA(2) \bbphi(3)+ \bbA(1)\bbphi(2)\bbphi(3) 
\bigr) \nn\\
&& e^{\t_0\sum_{r=1}^3 \frac{1}{\a_r} \left(\frac{(p^{(r)})^2}{2}-1 \right)} \left(\sum_{r=1}^3\bar N^r_1 a^{(r)\dag}_{1} \cdot \bbP\right) \vert 0 \rangle , \\
{\cal I}_{\phi A A} &=& \frac{2g}{3}\int \prod_{r=1}^3 dp^{(r)} \d \left(\sum_{r=1}^3 p^{(r)}\right) \, 
\text{tr}\, \langle 0 \vert  \bigl(\bbphi(1)\bbA(2) \bbA(3) + \bbA(1)\bbphi(2) \bbA(3)+ \bbA(1)\bbA(2)\bbphi(3) 
\bigr)\nn\\
&& e^{\t_0\sum_{r=1}^3 \frac{1}{\a_r} \left(\frac{(p^{(r)})^2}{2}-1 \right)}\left\{\half \sum_{r,s=1}^3 \bar N^{rs}_{11} a^{(r)\dagger}_{1} \cdot a^{(s)\dagger}_{1} + 
\frac{1}{2!}\left(\sum_{r=1}^3\bar N^r_1 a^{(r)\dagger}_{1} \cdot \bbP \right)^2
\right\} \vert 0 \rangle, \\
{\cal I}_{AAA} &=& \frac{2g}{3}\int \prod_{r=1}^3 dp^{(r)} \d \left(\sum_{r=1}^3 p^{(r)}\right) \, 
\text{tr}\, \langle 0 \vert \bbA(1)\bbA(2) \bbA(3)
e^{\t_0\sum_{r=1}^3 \frac{1}{\a_r} \left(\frac{(p^{(r)})^2}{2}-1 \right)} \nn\\
&&  \left\{\half \sum_{r,s=1}^3 \bar N^{rs}_{11} a^{(r)\dagger}_{1} \cdot a^{(s)\dagger}_{1}
\left(\sum_{r=1}^3\bar N^r_1 a^{(r)\dagger}_{1} \cdot \bbP \right) + 
\frac{1}{3!}\left(\sum_{r=1}^3\bar N^r_1 a^{(r)\dagger}_{1} \cdot \bbP \right)^3
\right\} \vert 0 \rangle.
\eeq 
\end{subequations}
The second term in ${\cal I}_{\phi AA}$ and the second term in ${\cal I}_{AAA}$ correspond to alpha-prime 
corrections to three-particle interactions.
Here, we note that there exists a three-particle interaction of two gauge particles and one tachyon, which 
may generate a four-gauge particle scattering mediated by a tachyon.

Using algebra we obtain 
\beq
{\cal I}_{\phi\phi A} &=& \frac{2g}{3}\int \prod_{r=1}^3 dp^{(r)} \d \left(\sum_{r=1}^3 p^{(r)}\right) \, 
\text{tr}\, \langle 0 \vert \bigl( \frac{1}{2} \bbphi(1)\bbphi(2) \bbA(3) + 4\bbphi(1)\bbA(2) \bbphi(3)+ 4\bbA(1)\bbphi(2)\bbphi(3) 
\bigr) \nn\\
&& \left(\sum_{r=1}^3\bar N^r_1 a^{(r)\dag}_{1} \cdot \bbP\right) \vert 0 \rangle \nn\\
&=& g \int \prod_{r=1}^3 dp^{(r)} \d \left(\sum_{r=1}^3 p^{(r)}\right) \, 
\text{tr}\, \Bigl\{A(1)\cdot (p^{(2)}-p^{(3)}) \phi(2) \phi(3) \Bigr\} \nn\\
&=& g \int \prod_{r=1}^3 dp^{(r)} \d \left(\sum_{r=1}^3 p^{(r)}\right) \, \text{tr}\,
p^{(1)}_\m \phi(p{(1)}) \left[A^\m (p^{(2)}), \phi(p^{(3)})\right].
\eeq 
In configuration space, it contributes to the action as 
\beq 
{\cal I}_{\phi\phi A}&=& -g i\int d^d x \,\text{tr} \, \p_\m \phi \left[A^\m, \phi \right] . 
\eeq 
Similarly, we have 
\beq
{\cal I}_{\phi A A} &=& \frac{2g}{3}\int \prod_{r=1}^3 dp^{(r)} \d \left(\sum_{r=1}^3 p^{(r)}\right) \, 
\text{tr}\, \langle 0 \vert  \bigl(2\bbphi(1)\bbA(2) \bbA(3) + 2\bbA(1)\bbphi(2) \bbA(3)+ 2^4\bbA(1)\bbA(2)\bbphi(3) 
\bigr)\nn\\
&&\left\{\half \sum_{r,s=1}^3 \bar N^{rs}_{11} a^{(r)\dagger}_{1} \cdot a^{(s)\dagger}_{1} + 
\frac{1}{2!}\left(\sum_{r=1}^3\bar N^r_1 a^{(r)\dagger}_{1} \cdot \bbP \right)^2
\right\} \vert 0 \rangle \nn\\
&=& \frac{2g}{3}\int \prod_{r=1}^3 dp^{(r)} \d \left(\sum_{r=1}^3 p^{(r)}\right) \, 
\text{tr}\, \Biggl\{\frac{3}{2} \phi(1) A(2)\cdot A(3)- \frac{3}{2} \phi(1) A(2) \cdot p^{(3)} A(3) \cdot p^{(2)}
\Biggr\} \nn\\
&=& g \int d^d x \, \text{tr}\, \left(\phi A_\m A^\m + \phi\, \p_\m A^\n \p_\n A^\m \right).
\eeq

The three-gauge particle interaction may be written as 
\begin{subequations}
\beq
{\cal I}_{AAA} &=& {\cal I}_{AAA}^{(0)} + {\cal I}_{AAA}^{(1)}, \\
{\cal I}_{AAA}^{(0)} &=& \frac{2^4g}{3}\int \prod_{r=1}^3 dp^{(r)} \d \left(\sum_{r=1}^3 p^{(r)}\right) \, 
\text{tr}\, \langle 0 \vert \bbA(1)\bbA(2) \bbA(3) \nn\\
&&  \left\{\half \sum_{r,s=1}^3 \bar N^{rs}_{11} a^{(r)\dagger}_{1} \cdot a^{(s)\dagger}_{1}
\left(\sum_{r=1}^3\bar N^r_1 a^{(r)\dagger}_{1} \cdot \bbP \right)\right\}\vert 0 \rangle , \\
{\cal I}_{AAA}^{(1)} &=& \frac{2^4g}{3}\int \prod_{r=1}^3 dp^{(r)} \d \left(\sum_{r=1}^3 p^{(r)}\right) \, 
\text{tr}\, \langle 0 \vert \bbA(1)\bbA(2) \bbA(3) \nn\\
&&  \left\{
\frac{1}{3!}\left(\sum_{r=1}^3\bar N^r_1 a^{(r)\dagger}_{1} \cdot \bbP \right)^3
\right\} \vert 0 \rangle.
\eeq 
\end{subequations}
Using the Neumann functions for three-string scattering, we get
\begin{subequations} 
\beq
{\cal I}_{AAA}^{(0)} 
&=& g \int \prod_{i=1} dp^{(i)} \d \left(\sum_{i=1}^3 p^{(i)} \right) p^\mu_1 
\,\text{tr} \Bigl( A^\n(p_1)  \left[ A_\n(p_2), A_\m(p_3)\right] \Bigr), \\
{\cal I}_{AAA}^{(1)} 
&=& \frac{g}{3}  \int \prod_{i=1}^3 dp^{(i)} \d\left(\sum_{i=1}^3 p^{(i)} \right) \, \text{tr} \, 
\Biggl\{\left(A^{(1)} \cdot p^{(2)}\right) \left(A^{(2)} \cdot p^{(3)}\right) \left(A^{(3)} \cdot p^{(1)}\right) \nn\\
&& -\left(A^{(1)} \cdot p^{(3)}\right) \left(A^{(2)} \cdot p^{(1)}\right) \left(A^{(3)} \cdot p^{(2)}\right)
\Biggr\} .
\eeq 
\end{subequations}
In configuration space they may be represented by   
three-gauge-field interaction terms of the $U(N)$ non-Abelian group Yang-Mills gauge theory: 
\begin{subequations} 
\beq
{\cal I}_{AAA}^{(0)} 
&=& g \int d^d x ~ i\,\text{tr} \left(\p_\m A_\n - \p_\n A_\m \right) \left[ A^\m, A^\n \right] \label{gauge3},\\ 
{\cal I}_{AAA}^{(1)} 
&=& \frac{g}{3}i \int d^d x \, \text{tr}\, \left(\p_\m A^\n- \p^\n A_\m \right)
\left(\p_\n A^\l- \p^\l A_\n \right)\left(\p_\l A^\m- \p^\m A_\l \right).
\eeq
\end{subequations}
It is worth mentioning that ${\cal I}_{AAA}^{(1)}$ is completely consistent with the $\ap$-correction to the three-gauge field interaction term, which has been obtained by previous
approaches \cite{Neveu72,Scherk74,Tseytlin86,Coletti03}.

\section{Four-gauge-particle scattering amplitude on space-filling Branes} 

The four-gauge-particle scattering amplitude has been discussed previously in Refs. \cite{Lee2016i,TLee2017cov} in the framework of string field theory in the proper-time gauge. However, in the previous works, we only studied the four-gauge-particle scattering amplitude 
in the zero-slope limit. Here, we shall evaluate the amplitude without taking the limit so that the resultant amplitude is valid for the full range of the energy scale. By mapping the string world-sheet for the four-string scattering onto the upper half complex plane by the Schwarz-Christoffel function, we may find that the four-string scattering amplitude on multiple space-filling branes may be written at tree level \cite{Lee2016i} as

\begin{subequations} 
\beq 
{\cal I}_{[4]} &=& 2g^2 \int \left\vert\frac{\prod_{r=1}^4 dZ_r }{ dV_{abc}}\right\vert 
\prod_{r<s} \vert Z_r - Z_s \vert^{p_r \cdot p_s} \exp\left[-\sum_{r=1}^4 \bar N^{[4]rr}_{00} \right]\nn\\
&& ~~~~~~\text{tr}\,\bigl\langle \Psi^{(1)}, \Psi^{(2)}, \Psi^{(3)}, \Psi^{(4)} \bigl\vert \exp \left[E_{[4]} \right] \bigr\vert 0 \bigr\rangle , \label{4scattering1}\\
E_{[4]} &=&   \half \sum_{r,s=1}^4 \sum_{m, n > 0} 
\bar N^{[4] rs}_{mn}\a^{(r)\dag}_{m\m} \a^{(s)\dag}_{n\n}\eta^{\m\n} +
\sum_{r,s=1}^4 \sum_{n > 0} 
\bar N^{[4] rs}_{n0}\a^{(r)\dag}_{n\m} p^{(s)\dag}_{\n}\eta^{\m\n} 
\label{4scattering2}
\eeq
\end{subequations}
where $Z_r$ $r=1,2,3,4$ denote the Koba-Nelson variables, corresponding to the locations of four external strings on upper half complex plane,  
\beq
Z_1 = 0, ~~Z_2 = x, ~~ Z_3 = 1, ~~ Z_4 = \infty. 
\eeq 
In the proper-time gauge, we choose $\a_1 =1, ~~~ \a_2 =1, ~~~ \a_3 =-1, ~~~ \a_4 =-1$.

In order to evaluate the four-gauge particle scattering amplitude, we choose the external string states as 
\beq \label{4states}
\Bigl\langle \Psi^{(1)}, \Psi^{(2)}, \Psi^{(3)}, \Psi^{(4)} \Bigl\vert  = \Bigl\langle 0 \Bigl\vert \prod_{r=1}^4\Bigl\{A_\m(p^{(r)}) a^{(r)}_{1\n} \eta^{\m\n} \Bigr\}.
\eeq  
Expanding the four-string vertex operator $E_{[4]}$ in terms of oscillator operators, 
we may find that the four-gauge particle scattering amplitude is given by 
\beq
{\cal I}_{AAAA} &=& 2g^2 \int \left\vert\frac{\prod_{r=1}^4 dZ_r }{ dV_{abc}}\right\vert 
\prod_{r<s} \vert Z_r - Z_s \vert^{p_r \cdot p_s} \exp\left[-\sum_{r=1}^4 \bar N^{[4]rr}_{00} \right]\text{tr}\, \langle 0 \vert \bbA(1) \bbA(2) \bbA(3) \bbA(4) \nn\\
&&  \Biggl\{\frac{1}{2!} \left(\half \sum_{r,s=1}^4 \bar N^{[4]rs}_{11} a^{(r)\dagger}_{1} \cdot a^{(s)\dagger}_{1}\right)^2 + 
\left(\half \sum_{r,s=1}^4 \bar N^{[4]rs}_{11} a^{(r)\dagger}_{1} \cdot a^{(s)\dagger}_{1}\right)\frac{1}{2!}\left(\sum_{r,s=1}^4\bar N^{[4]rs}_{10} a^{(r)\dagger}_{1} \cdot p^{(s)} \right)^2\nn\\
&&+ \frac{1}{4!}\left(\sum_{r,s=1}^4\bar N^{[4]rs}_{10} a^{(r)\dagger}_{1} \cdot p^{(s)} \right)^4
\Biggr\} \vert 0 \rangle.
\eeq 
If we explicitly reintroduce $\ap$, the momenta $p^{(r)}$ may be replaced by $\sqrt{\ap}p^{(r)}$, and this expansion can also be understood as a series expansion of the four-gauge particle scattering amplitude in powers of $\ap$. 
The third term in ${\cal I}_{AAAA}$ yields the alpha prime correction, which would have been missed if we had employed the vertex operator technique. It may be convenient to separately calculate each of the three terms in ${\cal I}_{AAAA}$.

\begin{figure}[htbp]
   \begin {center}
    \epsfxsize=0.9\hsize

	\epsfbox{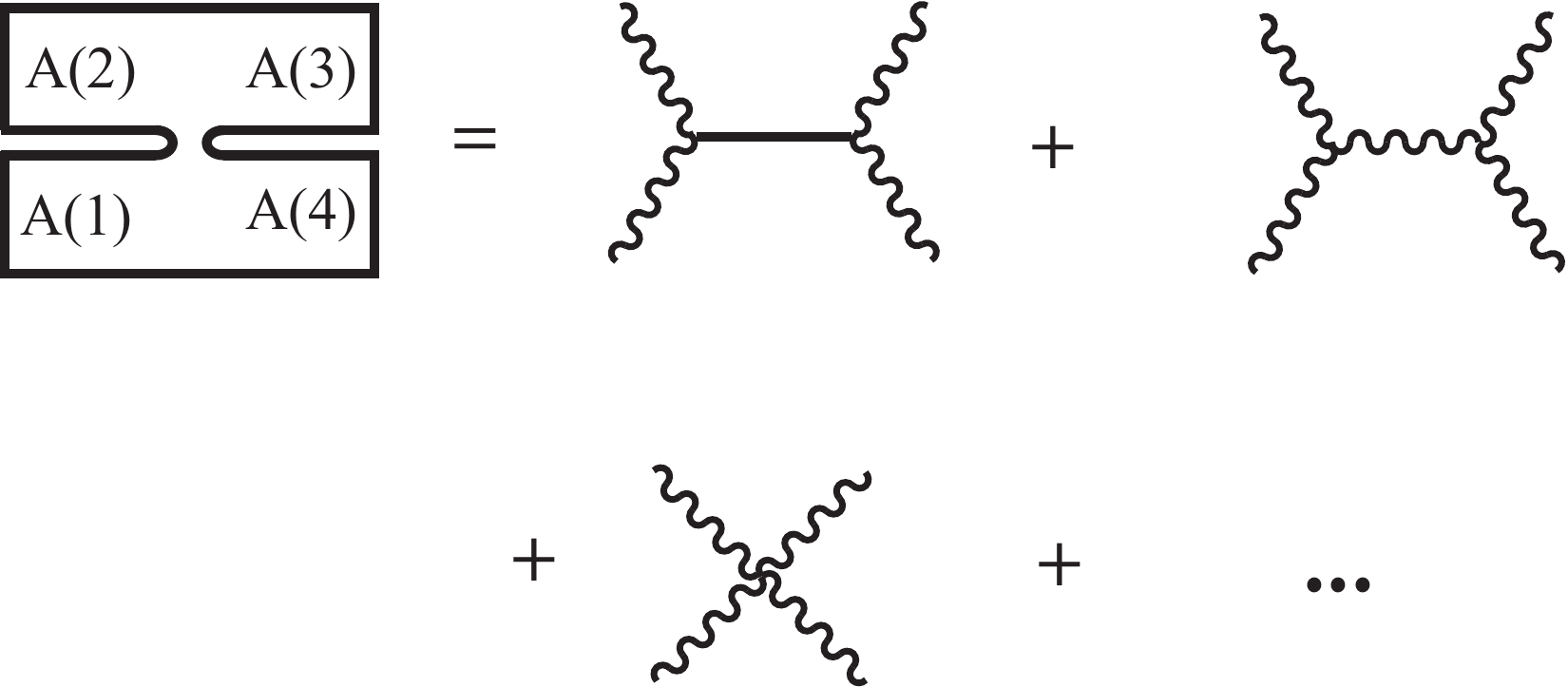}
   \end {center}
   \caption {\label{fourgauge} Four-gauge particle scattering amplitudes in open string theory.}
\end{figure}

\begin{figure}[htbp]
   \begin {center}
    \epsfxsize=1.0\hsize

	\epsfbox{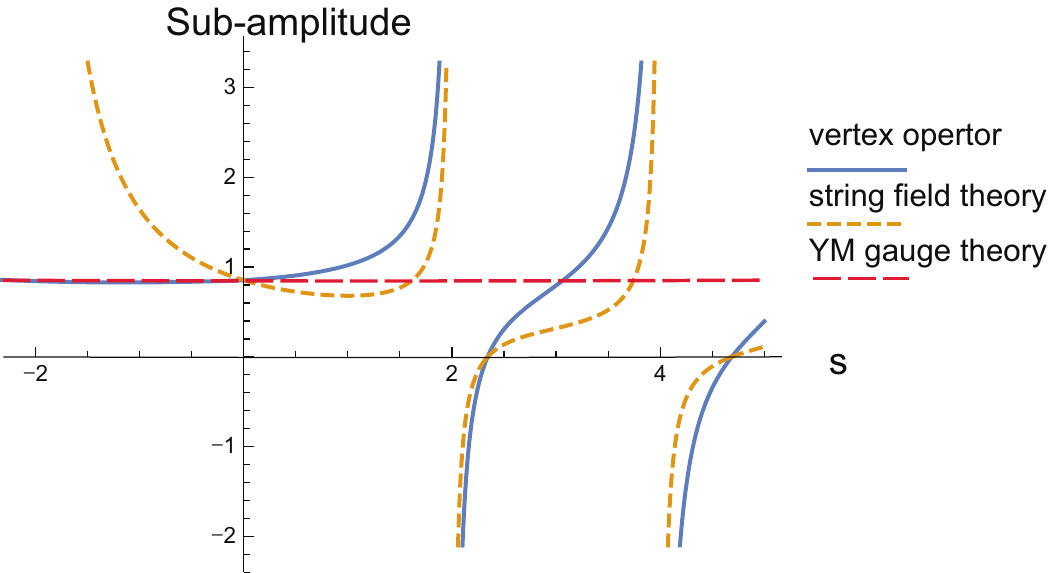}
   \end {center}
   \caption {\label{subamp} Sub-amplitudes of ${\cal I}^{(0)}$ in $s$-channel at a fixed angle.}
\end{figure}

The first term in ${\cal I}_{AAAA}$ may be defined by
\beq
{\cal I}_{AAAA}^{(0)} &=& 2g^2 \int \left\vert\frac{\prod_{r=1}^4 dZ_r }{ dV_{abc}}\right\vert 
\prod_{r<s} \vert Z_r - Z_s \vert^{p_r \cdot p_s} \exp\left[-\sum_{r=1}^4 \bar N^{[4]rr}_{00} \right]\text{tr}\, \langle 0 \vert \bbA(1) \bbA(2) \bbA(3) \bbA(4) \nn\\
&& \frac{1}{2!} \left(\half \sum_{r,s=1}^4 \bar N^{[4]rs}_{11} a^{(r)\dagger}_{1} \cdot a^{(s)\dagger}_{1}\right)^2
\vert 0 \rangle .
\eeq 
Using the Neumann functions given in the Appendix, we may evaluate ${\cal I}_{AAAA}^{(0)}$ as follows:
\beq \label{AAAA0string}
{\cal I}_{AAAA}^{(0)} &=& {\rm tr} \Bigl\{ A(1)_{\m_1} A(2)_{\m_2} A(3)_{\m_3} A(4)_{\m_4}
\Bigr\}\frac{\G\left(-\ap \frac{s}{2} \right) \G\left(-\ap \frac{t}{2} \right)}{\G\left(\ap \frac{u}{2}+1 \right)} \nn\\
&& \ap^2\Biggl\{\frac{tu}{4} \frac{1}{\ap s/2+1} \eta^{\m_1 \m_2} \eta^{\m_3 \m_4} + 
\frac{st}{4} \frac{1}{\ap u/2+1} \eta^{\m_1 \m_3} \eta^{\m_2 \m_4} +
\frac{us}{4} \frac{1}{\ap t/2+1}\eta^{\m_1 \m_4} \eta^{\m_2 \m_3} \Biggr\}
\eeq 
where $\ap$ is restored and $A(p^{(r)})$, $r=1,2,3,4$ are abbreviated as $A^a(r) T^a$, $r=1,2,3,4$. 
It can be compared with the corresponding sub-amplitude \cite{Schwarz1982} obtained by using the vertex operator technique
\beq \label{AAAA0schwarz}
{\cal I}_{AAAA}^{\rm vertex(0)} &=& {\rm tr} \Bigl\{ A(1)_{\m_1} A(2)_{\m_2} A(3)_{\m_3} A(4)_{\m_4}
\Bigr\}\frac{\G\left(-\ap\frac{s}{2} \right) \G\left(-\ap\frac{t}{2} \right)}{\G\left(\ap\frac{u}{2}+1 \right)} \nn\\
&&\ap^2\Biggl\{\frac{tu}{4} \eta^{\m_1 \m_2} \eta^{\m_3 \m_4} + 
\frac{st}{4} \eta^{\m_1 \m_3} \eta^{\m_2 \m_4} +
\frac{us}{4} \eta^{\m_1 \m_4} \eta^{\m_2 \m_3} \Biggr\}.
\eeq 
In the zero slope limit, both scattering amplitudes reduce to the corresponding sub-amplitude of non-Abelian 
gauge theory
\beq
{\cal I}_{AAAA}^{\rm YM(0)} &=& {\rm tr} \Bigl\{ A(1)_{\m_1} A(2)_{\m_2} A(3)_{\m_3} A(4)_{\m_4}
\Bigr\} \frac{4}{st} \nn\\
&&\Biggl\{\frac{tu}{4} \eta^{\m_1 \m_2} \eta^{\m_3 \m_4} + 
\frac{st}{4} \eta^{\m_1 \m_3} \eta^{\m_2 \m_4} +
\frac{us}{4} \eta^{\m_1 \m_4} \eta^{\m_2 \m_3} \Biggr\}.
\eeq 
However, we also notice the difference between two amplitudes: ${\cal I}_{AAAA}^{\rm vertex(0)}$ does not contain the tachyon poles whereas ${\cal I}_{AAAA}^{(0)}$ has the tachyon poles in all three channels: They differ for any finite $\alpha'$ and the amplitudes of Eq. (\ref{AAAA0string}) and Eq. (\ref{AAAA0schwarz}) only coincide 
at $\ap=0$. As we have observed in the last section, the three-open-string vertex gives rise to various three-particle interaction terms, including a coupling between two-gauge field and one tachyon when external string states are expanded in terms of mass eigenstates. Thus, the four-string scattering amplitude which is generated by the three-string vertex
should contain the tachyon poles. Fig. \ref{subamp} depicts the difference between the sub-amplitudes 
${\cal I}_{AAAA}^{(0)}$ in $s$-channel at a fixed angle. Thanks to the tachyon pole, the sub-amplitude of four-gauge particle scattering amplitude in $s$-channel changes significantly with finite $\ap$.

The second term in ${\cal I}_{AAAA}$ which is of order $\ap$ is written as 
\beq
{\cal I}_{AAAA}^{(1)} &=& 2g^2 \int \left\vert\frac{\prod_{r=1}^4 dZ_r }{ dV_{abc}}\right\vert 
\prod_{r<s} \vert Z_r - Z_s \vert^{p_r \cdot p_s} \exp\left[-\sum_{r=1}^4 \bar N^{[4]rr}_{00} \right]\text{tr}\, \langle 0 \vert \bbA(1) \bbA(2) \bbA(3) \bbA(4) \nn\\
&&\left(\half \sum_{r,s=1}^4 \bar N^{[4]rs}_{11} a^{(r)\dagger}_{1} \cdot a^{(s)\dagger}_{1}\right)\frac{1}{2!}\left(\sum_{r,s=1}^4\bar N^{[4]rs}_{10} a^{(r)\dagger}_{1} \cdot p^{(s)} \right)^2 \vert 0 \rangle .
\eeq 
Using the Neumann functions in the proper-time gauge given in the Appendix, we find that 
${\cal I}_{AAAA}^{(1)}$ may be expressed as an integral over the real Koba-Nelson variable $x$, 
\beq
{\cal I}_{AAAA}^{(1)} 
&=& {\rm tr} \Bigl\{ A_{\m_1}(1) A_{\m_2}(2) A_{\m_3}(3) A_{\m_4}(4)
\Bigr\} \int_0^1 dx \,\, x^{- \frac{s}{2}} (1-x)^{- \frac{t}{2}}
\nn\\
&&  
\Biggl\{- \eta^{\m_1 \m_2} \frac{1}{4} \frac{1}{(1-x)x^{2}} \left(x p^{(1)\m_3} + p^{(4)\m_3}\right)
\left(x p^{(2)\m_4} + p^{(3)\m_4}\right) \nn\\
&& + \eta^{\m_1 \m_3} \frac{1}{4} \frac{1}{(1-x)x} \left(p^{(1)\m_2} + x p^{(4)\m_2}\right)
\left(x p^{(2)\m_4} + p^{(3)\m_4}\right) \nn\\
&& - \eta^{\m_1 \m_4} \frac{1}{4} \frac{1}{(1-x)^2 x} \left((1-x)p^{(4)\m_2} + p^{(3)\m_2}\right)
\left((1-x) p^{(1)\m_3} + p^{(2)\m_3}\right) \nn\\
&& - \eta^{\m_2 \m_3} \frac{1}{4} \frac{1}{(1-x)^2 x} \left((1-x) p^{(3)\m_1} + p^{(4)\m_1}\right)
\left(p^{(1)\m_4} + (1-x) p^{(2)\m_4} \right) \nn\\
&& + \eta^{\m_2 \m_4} \frac{1}{4} \frac{1}{(1-x) x} \left(x p^{(3)\m_1} + p^{(2)\m_1}\right)
\left(x p^{(1)\m_3} + p^{(4)\m_3}\right) \nn\\
&& - \eta^{\m_3 \m_4} \frac{1}{4} \frac{1}{(1-x)x^{2}} \left(x p^{(3)\m_1} + p^{(2)\m_1}\right)
\left(p^{(1)\m_2} + x p^{(4)\m_2}\right) 
\Biggr\}. 
\eeq 
Integrating out the Koba-Nielsen variable $x$ leads us to 
\beq
{\cal I}_{AAAA}^{(1)} 
&=& {\rm tr}  \Bigl\{ A_{\m_1}(1) A_{\m_2}(2) A_{\m_3}(3) A_{\m_4}(4)\Bigr\}
\frac{\G\left(-\frac{s}{2} \right) \G\left(-\frac{t}{2} \right)}{\G\left(\frac{u}{2}+1 \right)} \nn\\
&&\Biggl\{- \frac{1}{4}\eta^{\m_1 \m_2} \left(-\frac{s}{2}p^{(1)\m_3} p^{(2)\m_4}+ 
\frac{u}{2} p^{(1)\m_3} p^{(3)\m_4}+ \frac{u}{2} p^{(4)\m_3} p^{(2)\m_4}
- \frac{u(u-2)}{4} \frac{1}{\frac{s}{2}+1} p^{(4)\m_3} p^{(3)\m_4}\right)  \nn\\
&&+ \frac{1}{4}\eta^{\m_1 \m_3}\left( 
-\frac{s}{2} p^{(1)\m_2} p^{(2)\m_4}+ \frac{u}{2} p^{(1)\m_2} p^{(3)\m_4}+ 
\frac{s(s-2)}{4} \frac{1}{\frac{u}{2}+1} p^{(4)\m_2} p^{(2)\m_4}- \frac{s}{2} p^{(4)\m_2}p^{(3)\m_4}\right)\nn\\
&&- \frac{1}{4}\eta^{\m_1 \m_4} \left(-\frac{t}{2} p^{(4)\m_2} p^{(1)\m_3}+ \frac{u}{2} p^{(4)\m_2} p^{(2)\m_3}+ \frac{u}{2} p^{(3)\m_2} p^{(1)\m_3}- \frac{u(u-2)}{4} \frac{1}{\frac{t}{2}+1} p^{(3)\m_2} p^{(2)\m_3}\right)\nn\\
&&- \frac{1}{4}\eta^{\m_2 \m_3} \left(-\frac{t}{2} p^{(3)\m_1} p^{(2)\m_4}+ \frac{u}{2} p^{(3)\m_1} p^{(1)\m_4}+ \frac{u}{2} p^{(4)\m_1} p^{(2)\m_4}- \frac{u(u-2)}{4} \frac{1}{\frac{t}{2}+1} p^{(4)\m_1} p^{(1)\m_4}\right)\nn\\
&&+ \frac{1}{4}\eta^{\m_2 \m_4} \left(\frac{s(s-2)}{4} \frac{1}{\frac{u}{2}+1} p^{(3)\m_1} p^{(1)\m_3}
-\frac{s}{2} p^{(3)\m_1} p^{(4)\m_3} -\frac{s}{2} p^{(2)\m_1} p^{(1)\m_3}+\frac{u}{2} p^{(2)\m_1} p^{(4)\m_3}\right)\nn\\
&&- \frac{1}{4}\eta^{\m_3 \m_4} \left(-\frac{s}{2} p^{(3)\m_1} p^{(4)\m_2}+ \frac{u}{2} p^{(3)\m_1} p^{(1)\m_2}+ \frac{u}{2} p^{(2)\m_1} p^{(4)\m_2} - \frac{u(u-2)}{4} \frac{1}{\frac{s}{2}+1} p^{(2)\m_1} p^{(1)\m_2}\right)
\Biggr\}.
\eeq
Again, we find that the scattering amplitude ${\cal I}_{AAAA}^{(1)}$ contains the tachyon poles in all three channels. It may be interesting to compare this sub-amplitude with the corresponding one, which can
be obtained using the conventional vertex operator technique. If we apply the vertex operator technique to evaluate the corresponding sub-amplitude, we obtain \cite{Schwarz1982}
\beq
{\cal I}_{AAAA}^{\rm vertex(1)} &=& {\rm tr}  \Bigl\{ A_{\m_1}(1) A_{\m_2}(2) A_{\m_3}(3) A_{\m_4}(4)\Bigr\}
\frac{\G\left(-\frac{s}{2} \right) \G\left(-\frac{t}{2} \right)}{\G\left(\frac{u}{2}+1 \right)} \nn\\
&&\Bigg\{\frac{\eta^{\m_1 \m_2}}{2} \left(t p^{(1)\m_3} p^{(2)\m_4}+ u p^{(2)\m_3} p^{(1)\m_4} \right) + \frac{\eta^{\m_1 \m_3}}{2} \left(t p^{(1)\m_2} p^{(3)\m_4}+ s p^{(3)\m_2} p^{(1)\m_4} \right)  \nn\\
&& + \frac{\eta^{\m_1 \m_4}}{2} \left(u p^{(1)\m_2} p^{(4)\m_3}+ s p^{(4)\m_2} p^{(1)\m_3} \right) + \frac{\eta^{\m_2 \m_3}}{2} \left(u p^{(2)\m_1} p^{(3)\m_4}+ s p^{(3)\m_1} p^{(2)\m_4}  \right)  \nn\\
&& + \frac{\eta^{\m_2 \m_4}}{2} \left(t p^{(2)\m_1} p^{(4)\m_3}+s p^{(4)\m_1} p^{(2)\m_3}  \right) + \frac{\eta^{\m_3 \m_4}}{2} \left(t p^{(3)\m_1} p^{(4)\m_2}+ u p^{(4)\m_1} p^{(3)\m_2}  \right) \Biggr\}.
\eeq 
The main difference between two sub-amplitudes ${\cal I}_{AAAA}^{(1)}$ and ${\cal I}_{AAAA}^{\rm vertex(1)}$ is the presence of tachyon poles.  
In the zero-slope limit, both scattering amplitudes reduce to the corresponding one of Yang-Mills gauge theory
as expected
\beq
{\cal I}_{AAAA}^{\rm YM (1)} &=& {\rm tr}  \Bigl\{ A_{\m_1}(1) A_{\m_2}(2) A_{\m_3}(3) A_{\m_4}(4)\Bigr\}
\frac{4}{st} \nn\\
&&\Bigg\{\frac{\eta^{\m_1 \m_2}}{2} \left(t p^{(1)\m_3} p^{(2)\m_4}+ u p^{(2)\m_3} p^{(1)\m_4} \right) + \frac{\eta^{\m_1 \m_3}}{2} \left(t p^{(1)\m_2} p^{(3)\m_4}+ s p^{(3)\m_2} p^{(1)\m_4} \right)  \nn\\
&& + \frac{\eta^{\m_1 \m_4}}{2} \left(u p^{(1)\m_2} p^{(4)\m_3}+ s p^{(4)\m_2} p^{(1)\m_3} \right) + \frac{\eta^{\m_2 \m_3}}{2} \left(u p^{(2)\m_1} p^{(3)\m_4}+ s p^{(3)\m_1} p^{(2)\m_4}  \right)  \nn\\
&& + \frac{\eta^{\m_2 \m_4}}{2} \left(t p^{(2)\m_1} p^{(4)\m_3}+s p^{(4)\m_1} p^{(2)\m_3}  \right) + \frac{\eta^{\m_3 \m_4}}{2} \left(t p^{(3)\m_1} p^{(4)\m_2}+ u p^{(4)\m_1} p^{(3)\m_2}  \right) \Biggr\}.
\eeq

\section{The alpha prime corrections}

One of outstanding advantages of the string field theory in the proper-time gauge is that we may be able 
to systematically calculate the $\ap$-corrections . An expansion of $N$-string vertex operator $\exp[E_{[N]}]$ in terms of oscillator operators naturally yields a series expansion of $\ap$ with a unique ordering of non-Abelian operators. When we expand the vertex operator $\exp[E_{[4]}]$ 
Eq. (\ref{4scattering1}) in terms of oscillator operators, we find that the third term is proportional to $\ap^2$. 
This term may be considered as $\ap$-corrections to the four-gauge-particle scattering amplitude, which do not 
have counterparts in the conventional calculation using vertex operators: 
\beq
{\cal I}_{AAAA}^{(2)} &=& 2g^2 \int \left\vert\frac{\prod_{r=1}^4 dZ_r }{ dV_{abc}}\right\vert 
\prod_{r<s} \vert Z_r - Z_s \vert^{p_r \cdot p_s} \exp\left[-\sum_{r=1}^4 \bar N^{[4]rr}_{00} \right]\text{tr}\, \langle 0 \vert \bbA(1) \bbA(2) \bbA(3) \bbA(4) \nn\\
&& ~~\frac{1}{4!}\left(\sum_{r,s=1}^4\bar N^{[4]rs}_{10} a^{(r)\dagger}_{1} \cdot p^{(s)} \right)^4
\Biggr\} \vert 0 \rangle \nn\\
&=& 2g^2 \int^1_0 \,dx \, x^{-\frac{s}{2}+2} (1-x)^{-\frac{t}{2}-2} \,e^{2(\t_2-\t_1)}
\prod_{r=1}^4 \left\{\sum_{s=0}^4 \bar N^{[4]rs}_{10} A(r) \cdot p^{(s)} \right\}.
\eeq
Using the Neumann functions $\bar N^{[4]rs}_{10}$ evaluated in the Appendix and integrating out the real Koba-Nielsen variable $x$, we find that 
\beq
{\cal I}_{AAAA}^{(2)}
&=& (2g^2) {\rm tr} \Biggl\{A(1)^{\m_1} A(2)^{\m_2} A(3)^{\m_3} A(4)^{\m_4} \Biggr\} \frac{\G\left(-\frac{s}{2}\right) \G\left(-\frac{t}{2}\right)}{\G\left(\frac{u}{2}+1\right)} \nn\\
&& 
\frac{1}{\left(\frac{u}{2}+1\right)}
\Biggl\{ \frac{\frac{t}{2}\left(\frac{t}{2}-1\right)\left(\frac{t}{2}-2\right)}{\left(\frac{s}{2}+1\right)
} p^{(2)}_{\m_1}p^{(1)}_{\m_2}p^{(4)}_{\m_3}p^{(3)}_{\m_4} \nn\\
&& - \frac{t}{2}\left(\frac{t}{2}-1\right)\Bigl(
p^{(2)}_{\m_1}p^{(1)}_{\m_2}p^{(4)}_{\m_3}p^{(1)}_{\m_4}+  p^{(2)}_{\m_1}p^{(1)}_{\m_2}p^{(2)}_{\m_3}p^{(3)}_{\m_4}+
p^{(2)}_{\m_1}p^{(3)}_{\m_2}p^{(4)}_{\m_3}p^{(3)}_{\m_4}+
p^{(4)}_{\m_1}p^{(1)}_{\m_2}p^{(4)}_{\m_3}p^{(3)}_{\m_4}\Bigr) \nn\\
&& + \frac{st}{4}\Biggl(
p^{(2)}_{\m_1}p^{(1)}_{\m_2}p^{(2)}_{\m_3}p^{(1)}_{\m_4}+ 
p^{(2)}_{\m_1}p^{(3)}_{\m_2}p^{(4)}_{\m_3}p^{(1)}_{\m_4}+ 
p^{(2)}_{\m_1}p^{(3)}_{\m_2}p^{(2)}_{\m_3}p^{(3)}_{\m_4} \nn\\
&& + p^{(4)}_{\m_1}p^{(1)}_{\m_2}p^{(4)}_{\m_3}p^{(1)}_{\m_4}+ p^{(4)}_{\m_1}p^{(1)}_{\m_2}p^{(2)}_{\m_3}p^{(3)}_{\m_4}+ 
p^{(4)}_{\m_1}p^{(3)}_{\m_2}p^{(4)}_{\m_3}p^{(3)}_{\m_4}+ 
\Biggr)\nn\\
&&- \frac{s}{2}\left(\frac{s}{2}-1\right) 
\Bigl(p^{(2)}_{\m_1}p^{(3)}_{\m_2}p^{(2)}_{\m_3}p^{(1)}_{\m_4}+  p^{(4)}_{\m_1}p^{(1)}_{\m_2}p^{(2)}_{\m_3}p^{(1)}_{\m_4}+
p^{(4)}_{\m_1}p^{(3)}_{\m_2}p^{(4)}_{\m_3}p^{(1)}_{\m_4}+
p^{(4)}_{\m_1}p^{(3)}_{\m_2}p^{(2)}_{\m_3}p^{(3)}_{\m_4}\Bigr) \nn\\
&&+ \frac{\frac{s}{2}\left(\frac{s}{2}-1\right)\left(\frac{s}{2}-2\right)}{\left(\frac{t}{2}+1\right)
} p^{(4)}_{\m_1}p^{(3)}_{\m_2}p^{(2)}_{\m_3}p^{(1)}_{\m_4} 
\Biggr\}.
\eeq
It is apparent that the sub-amplitude ${\cal I}_{AAAA}^{(2)}$ also contains the tachyon poles.   
It would have been difficult to obtain these $\ap$-corrections to the four-gauge particle scattering amplitude if 
we employ the vertex opertor technique. In the zero slope limit, ${\cal I}_{AAAA}^{(2)}$ reduces to 
\beq
{\cal I}_{AAAA}^{(2)} &\rightarrow& (2g^2) 2 {\rm tr} \Biggl\{A(1)^{\m_1} A(2)^{\m_2} A(3)^{\m_3} A(4)^{\m_4} \Biggr\} \Biggl\{\frac{1}{s} \Bigl( 2 p^{(2)}_{\m_1}p^{(1)}_{\m_2}p^{(4)}_{\m_3}p^{(3)}_{\m_4} \nn\\
&& +
p^{(2)}_{\m_1}p^{(1)}_{\m_2}p^{(4)}_{\m_3}p^{(1)}_{\m_4}+  p^{(2)}_{\m_1}p^{(1)}_{\m_2}p^{(2)}_{\m_3}p^{(3)}_{\m_4}+
p^{(2)}_{\m_1}p^{(3)}_{\m_2}p^{(4)}_{\m_3}p^{(3)}_{\m_4}+
p^{(4)}_{\m_1}p^{(1)}_{\m_2}p^{(4)}_{\m_3}p^{(3)}_{\m_4}\Bigr) \nn\\
&& + \frac{1}{2}\Biggl(
p^{(2)}_{\m_1}p^{(1)}_{\m_2}p^{(2)}_{\m_3}p^{(1)}_{\m_4}+ 
p^{(2)}_{\m_1}p^{(3)}_{\m_2}p^{(4)}_{\m_3}p^{(1)}_{\m_4}+ 
p^{(2)}_{\m_1}p^{(3)}_{\m_2}p^{(2)}_{\m_3}p^{(3)}_{\m_4} \nn\\
&& + p^{(4)}_{\m_1}p^{(1)}_{\m_2}p^{(4)}_{\m_3}p^{(1)}_{\m_4}+ p^{(4)}_{\m_1}p^{(1)}_{\m_2}p^{(2)}_{\m_3}p^{(3)}_{\m_4}+ 
p^{(4)}_{\m_1}p^{(3)}_{\m_2}p^{(4)}_{\m_3}p^{(3)}_{\m_4}
\Biggr)\nn\\
&& +\frac{1}{t}
\Bigl(p^{(2)}_{\m_1}p^{(3)}_{\m_2}p^{(2)}_{\m_3}p^{(1)}_{\m_4}+  p^{(4)}_{\m_1}p^{(1)}_{\m_2}p^{(2)}_{\m_3}p^{(1)}_{\m_4}+
p^{(4)}_{\m_1}p^{(3)}_{\m_2}p^{(4)}_{\m_3}p^{(1)}_{\m_4}+
p^{(4)}_{\m_1}p^{(3)}_{\m_2}p^{(2)}_{\m_3}p^{(3)}_{\m_4} \nn\\
&&+ 2 p^{(4)}_{\m_1}p^{(3)}_{\m_2}p^{(2)}_{\m_3}p^{(1)}_{\m_4}\Bigr) 
\Biggr\}.
\eeq  
From this expression of ${\cal I}_{AAAA}^{(2)}$, we may infer that the 
corresponding $\ap$-corrections to the non-Abelian Yang-Mills action are of order $F^4$.

\section{Discussions and Conclusions}

We conclude this work with a few remarks on possible extensions. We calculated the four-gauge-particle treee level scattering amplitudes on multiple space-filling D-branes using the Polyakov string path integral in the proper-time gauge. Although the resultant scattering amplitudes are 
completely consistent with the conventional ones obtained by substituting local vertex operators for external strings in the low energy region, they significantly differ from the conventional ones when string corrections become relevant with finite $\ap$, as they contain tachyon poles in all three scattering channels in a manner consistent with three-particle interactions and $\ap$-corrections, which are of order $F^4$.

The string scale may be as low as the weak scale \cite{Lykken1996,Antoniadis1998} in some phenomenological models based on string theory, so it is important to accurately evaluate 
particle scattering amplitudes, which are valid for the full range of the energy scale. 
The Polyakov string path integral in the proper-time gauge, which is equivalent to the deformed cubic string field theory \cite{Lee2017d} may be the right theoretical tool to handle this request. The momenta of external strings are not restricted to the low energy region and the multi-particle scattering amplitudes evaluated in the proper-time gauge yield series expansions of $\ap$ with an unambiguously defined ordering of non-Abelian field operators.

Particle scattering amplitudes satisfy various relationships between themselves such as the Kleiss-Kuijf (KK) relation \cite{Kleiss1989} and the Bern-Carrasco-Johansson (BCJ) relation \cite{BCJ}, and there has been much effort toward understanding the origins of these relations in string theory \cite{Chan2006notes,Stieberger09,Bjerrum-Bohr2014,Stieberger2016,Lai2016string}. Because the multi-particle scattering amplitudes obtained by evaluating Polyakov string path integral in the proper-time gauge can be interpreted as the Feynman-Schwinger proper-time representation of open string field theory, we may be able to study relations between scattering amplitudes of massive higher spin particles in string theory by extending this work. It may 
also be interesting to explore the scattering amplitudes of massless scalars and non-Abelian gauge fields by defining the scattering amplitudes of open strings on $Dp$-branes \cite{TLee2017cov}. It may also be worth noting
that the scattering amplitudes in the proper-time gauge are valid for the full range of energy scales and expanded in a power series of $\ap$. These properties make them useful tools for probing
high energy limits of string theory \cite{Gross87,Gross88,Gross88prl,Gross89Phil,Gross89nucl}.

\vskip 1cm

\begin{acknowledgments}
This work was supported by Basic Science Research Program through the National Research Foundation of Korea(NRF) funded by the Ministry of Education (2017R1D1A1A02017805). 
\end{acknowledgments}


\appendix

\section{Neumann Functions for Four-String Scattering Amplitudes}

In order to evaluate the Polakov string path integral, we map the world sheet of four-string scattering 
onto upper half complex plane: Four external strings are located on the boundary of upper half complex plane 
\beq
Z_1 = 0, ~~ Z_2 = x, ~~ Z_3 =1, ~~ Z_4 = \infty, ~~~ 0 \le x \le 1 .
\eeq 
Accordingly, the Schwarz-Christoffel transformation which maps the four-string-scattering world sheet onto 
upper half complex plane is constructed as 
\beq
\rho = \sum_{r=1}^4 \a_r \ln (z-Z_r) = \ln z + \ln (z-x) - \ln (1-z) = \ln \frac{z (z-x)}{(1-z)}  
\eeq 
where $\a_1 = \a_2 =1$, $\a_3=\a_4 =-1$. 
It follows from this mapping that relations between the global coordinate $\rho$ and the local coordinates $\zeta_i$, $i=1, 2, 3, 4$ on individual string patches as given by
\beq
e^{-\zeta_1} &=& e^{\t_1} \frac{(1-z)}{z(z-x)}, ~~~ e^{-\zeta_2} = -e^{\t_1} \frac{(1-z)}{z(z-x)}, \nn\\
e^{-\zeta_3} &=& e^{-\t_2} \frac{z(z-x)}{(1-z)}, ~~~ e^{-\zeta_4} = - e^{-\t_2} \frac{z(z-x)}{(1-z)} ,
\eeq 
where $\t_1$ and $\t_2$ are two intraction times on the world sheet. 

The Neumann functions $\bar N^{rs}_{n0}$ are given by contour integrals as follows 
\beq
\bar N^{rs}_{n0} &=& \frac{1}{n} \oint_{Z_r} \frac{dz}{2\pi i} \frac{1}{z- Z_s} e^{-n \zeta_r(z)}. 
\eeq 
Performing the contour integrals and the binormial series expansions we may 
explicitly evaluate $\bar N^{rs}_{n0}$
\beq
\bar N^{11}_{n0} 
= \frac{e^{n\t_1}}{n!}\frac{1}{x^n} \sum_{k=0}^{n} \left(\begin{array}{cc} n \\ k \end{array} \right)
\frac{(n+k-1)!}{k!}\left(\frac{-1}{x}\right)^k,
\eeq
\beq
\bar N^{12}_{n0} &=& 
=\frac{e^{n\t_1}}{n!}\frac{1}{x^{n+1}}\sum_{k=0}^{n-1} \left(\begin{array}{cc} n-1 \\ k \end{array} \right)
\frac{(n+k)!}{(k+1)!}\left(\frac{-1}{x}\right)^k, 
\eeq
\beq
\bar N^{13}_{n0} &=& \frac{e^{n\t_1}}{n!}\frac{1}{x^n} \sum_{k=0}^{n-1} \left(\begin{array}{cc} n-1 \\ k \end{array} \right)
\frac{(n+k-1)!}{k!}\left(\frac{-1}{x}\right)^k, 
\eeq
\beq
\bar N^{14}_{n0} &=& 0, \\
\bar N^{21}_{n0} 
&=& -\frac{e^{n\t_1}}{n!}\frac{1}{x^n} \sum_{k=0}^{n-1} \left(\begin{array}{cc} n-1 \\ k \end{array} \right)
\frac{(n+k)!}{(k+1)!}\left(\frac{1}{x}-1\right)^{k+1}, \\ 
\bar N^{22}_{n0}
&=& \frac{e^{n\t_1}}{n!}\frac{1}{x^n} \sum_{k=0}^{n} \left(\begin{array}{cc} n \\ k \end{array} \right)
\frac{(n+k-1)!}{k!}\left(\frac{1}{x}-1\right)^k, \\
\bar N^{23}_{n0} 
&=& \frac{e^{n\t_1}}{n!}\frac{1}{x^n} \sum_{k=0}^{n-1} \left(\begin{array}{cc} n-1 \\ k \end{array} \right)
\frac{(n+k-1)!}{k!}\left(\frac{1}{x}-1\right)^k, \\
\bar N^{24}_{n0} &=& 0,  \\
\bar N^{31}_{n0} 
&=& \frac{(-1)^n e^{-n\t_2}}{n}(1-x)^n \sum_{k=0}^{n-1} \left(\begin{array}{cc} n-1 \\ k \end{array} \right)
\left(\begin{array}{cc} n \\ k \end{array} \right) \frac{1}{(1-x)^k}, \\
\bar N^{32}_{n0} 
&=&\frac{(-1)^n e^{-n\t_2}}{n}(1-x)^{n} \sum_{k=0}^{n-1} \left(\begin{array}{cc} n-1 \\ k \end{array} \right)
\left(\begin{array}{cc} n \\ k+1 \end{array} \right) \frac{1}{(1-x)^{k+1}}, \\
\bar N^{33}_{n0}
&=& \frac{(-1)^n e^{-n\t_2}}{n} (1-x)^n \sum_{k=0}^{n} \left(\begin{array}{cc} n \\ k \end{array} \right)
\left(\begin{array}{cc} n \\ k \end{array} \right) \frac{1}{(1-x)^k}, \\
\bar N^{34}_{n0} &=& 0 , \\
\bar N^{41}_{n0}
&=&- \frac{(-1)^n e^{-n\t_2}}{n!} x^n \sum_{k=0}^{n} \left(\begin{array}{cc} n \\ k \end{array} \right)
\frac{(n+k-1)!}{k!} \left(\frac{-1}{x}\right)^k, \\
\bar N^{42}_{n0} 
&=&\frac{(-1)^n e^{-n\t_2}}{n} \frac{x^n}{n!} \sum_{k=1}^{n} \left(\begin{array}{cc} n \\ k \end{array} \right)
\frac{(n+k-1)!}{(k-1)!} \frac{1}{x^k}, \\
\bar N^{43}_{n0} 
&=&-\frac{(-1)^n e^{-n\t_2}}{n} \frac{x^n}{n!} \sum_{k=0}^{n} \left(\begin{array}{cc} n \\ k \end{array} \right)
\frac{(n+k)!}{k!}\left(\frac{-1}{x}\right)^k, \\
\bar N^{44}_{n0} &=& 0.
\eeq
For $n=1$, we have 
\beq
\bar N^{11}_{10} &=& - e^{\t_1} \frac{(1-x)}{x^2}, ~~~\bar N^{12}_{10} = \frac{e^{\t_1}}{x^2}, ~~~\bar N^{13}_{10} = \frac{e^{\t_1}}{x}, ~~~\bar N^{14}_{10} = 0,\\
\bar N^{21}_{10} &=& - e^{\t_1} \frac{(1-x)}{x^2}, ~~~\bar N^{22}_{10} =  \frac{ e^{\t_1}}{x^2}, ~~~
\bar N^{23}_{10} = \frac{e^{\t_1} }{x},~~~ \bar N^{24}_{10} =0,\\
\bar N^{31}_{10} &=& -e^{-\t_2} (1-x),~~~
\bar N^{32}_{10} = -e^{-\t_2}, ~~~ \bar N^{33}_{10} =  - e^{-\t_2} (2-x) , ~~~\bar N^{34}_{10} = 0 , \\
\bar N^{41}_{10} &=&- e^{-\t_2}(1-x) , ~~~\bar N^{42}_{10} = -e^{-\t_2} , ~~~\bar N^{43}_{10} = e^{-\t_2} (x-2) , ~~~
\bar N^{44}_{10} = 0.
\eeq

The Neuman functions $\bar N^{rs}_{nm}$ is defined by 
\beq
\bar N^{rs}_{nm} &=& \frac{1}{nm} \oint_{Z_r} \frac{dz}{2\pi i} \oint_{Z_s} \frac{d \zp}{2\pi i} 
\frac{1}{(z-\zp)^2} e^{-n\zeta_r(z) - m \zeta^\prime_s(\zp)}, ~~~ n, m \ge 1  
\eeq
We may explicitly calculate $\bar N^{rs}_{11}$, 
\beq
\bar N^{rs}_{11} &=& \oint_{Z_r} \frac{dz}{2\pi i} \oint_{Z_s} \frac{d \zp}{2\pi i} 
\frac{1}{(z-\zp)^2} e^{-\zeta_r(z) - \zeta^\prime_s(\zp)}, ~~~ \text{for}~~ r< s .
\eeq
Through algebra, we find that
\beq
\bar N^{12}_{11} &=& e^{2\t_1} \frac{(1-x)}{x^4},~~~~ 
\bar N^{13}_{11} = e^{\t_1- \t_2} \frac{(1-x)}{x},~~~~ 
\bar N^{14}_{11} = e^{\t_1 - \t_2} \frac{1}{x}, \\
\bar N^{23}_{11} &=& e^{\t_1- \t_2} \frac{1}{x},~~~~ 
\bar N^{24}_{11} = e^{\t_1- \t_2} \frac{(1-x)}{x},~~~~
\bar N^{34}_{11} = e^{-2\t_2} (1-x).
\eeq

\end{document}